\documentclass[aps,prl,twocolumn,floatfix]{revtex4}
\usepackage[latin1]{inputenc}
\usepackage{amsfonts}
\usepackage{amssymb}
\usepackage{amsmath}
\usepackage{calc}
\usepackage{graphicx}
\usepackage{epsfig}
\usepackage{bm}
\usepackage{ulem}
\usepackage{setspace}
\usepackage{subfigure}

\begin{document}

\title{Conscious observers clarify many worlds}
\date{\today}
\author{Christoph Simon}
\affiliation{Institute for Quantum Information Science and
Department of Physics and Astronomy, University of Calgary,
Calgary T2N 1N4, Alberta, Canada}

\begin{abstract}
In this brief note I argue that putting conscious observers
at the center of the considerations clarifies and
strengthens the many-worlds interpretation. The basic
assumption, which seems extremely plausible based on our
current understanding of the brain and of decoherence, is
that quantum states corresponding to distinct conscious
experiences have to be orthogonal. I show that, once this
is accepted, probabilistic measurement outcomes
corresponding to basis elements and following Born's rule
emerge naturally from global unitary dynamics.
\end{abstract}

\maketitle

Given the great success of quantum physics in the
microscopic domain, it is very natural to wonder whether it
is universally valid, and in particular whether it can
account for our macroscopic experience. Arguing that it can
is the program of the ``many worlds'' interpretation of
quantum physics, which was first proposed by Everett in
1957 \cite{Everett} and developed by many others since
then. In this brief note I argue that it is very useful in
the context of this approach to put ourselves, i.e.
conscious observers, at the center of the considerations.
This is not only natural, since our ultimate goal is to
explain our experience of the world (which at first sight
seems to be in contradiction with the quantum description),
but also has considerable explanatory power, especially
when combined with our modern understanding of decoherence
\cite{decoh-refs}. In particular, this approach helps to
clarify three points that are frequently raised in
discussions of ``many worlds'' \cite{zurekortho,kent}.
First, it explains why the results of ideal measurements
always correspond to bases in Hilbert space, even though
there is a continuum of possible quantum states ``between''
the basis elements. Second, it explains why we experience
definite outcomes even though the global quantum state is a
superposition of all possibilities. Third, it also takes us
a long way towards understanding why the probabilities of
these definite outcomes are given by the Born rule.

The reason why thinking about conscious observers is useful
in this context is the following. If we assume for the sake
of the argument that quantum physics applies to conscious
beings, then it is extremely plausible that {\it quantum
states that correspond to definite experiences} (e.g. the
experience of observing ``up'' or ``down'' in a
Stern-Gerlach type experiment) {\it are mutually
orthogonal}. This is plausible because neuroscience shows
already quite convincingly that different states of
conscious experience are associated with different patterns
of neural activity in the brain. Moreover the expected
decoherence times for superpositions of different firing
patterns are extremely short compared to the characteristic
timescales of conscious experience \cite{neuro-decoh}. It
therefore seems safe to assume that all the possible states
of experience of a conscious subject form part of a basis.
(Other elements of the same basis correspond to the
observer being unconscious, or dead.)

This means that if $|A_1\rangle$ and $|A_2\rangle$ are two
states of the conscious subject, Alice, corresponding to
two different states of her conscious experience, let's
call them ``experience 1'' and ``experience 2'', then
$\langle A_1|A_2\rangle=0$. Moreover the state $\alpha
|A_1\rangle + \beta |A_2\rangle$ (with $\alpha$ and $\beta$
both different from zero) does not correspond to any single
conscious experience. Instead it corresponds, from Alice's
point of view, to a superposition of two different
``worlds'', one in which she has experience 1, and one
where she has experience 2. Note that another observer,
Bob, may be perfectly aware that Alice is in a
superposition state. If he is technologically very
powerful, he could perform the appropriate quantum
interference experiment to prove it. In the course of such
an experiment Alice would have to lose any reliable memory
of her previous experience (i.e. whether it was experience
1 or experience 2). But there appears to be no fundamental
problem with such a scenario. In particular her conscious
experience does not constitute a hidden variable with
greater predictive power than the quantum state. On the
contrary, the quantum superposition state $\alpha
|A_1\rangle+ \beta |A_2\rangle$ tells the whole story about
her as a quantum system, whereas her experience (which
corresponds to $|A_1\rangle$ or $|A_2\rangle$, depending on
her ``world'') does not. One could even imagine Bob telling
Alice that really she is in a superposition state. She
would have to reply, ``This may be so, but based only on my
subjective experience there is no way for me to know.''

Let me now outline how this point of view helps to clarify
the many worlds approach. First, it explains why the
possible results of ideal measurements on a quantum system
always correspond to states in a certain orthogonal basis.
An ideal measurement of a quantum system in a state
$|S_1\rangle$ should always give the same result (``state
$|S_1\rangle$ !''). By the definition of measurement, this
result corresponds to a definite experience for the
observer. More formally, we can write the dynamics as
$|S_1\rangle|O_0\rangle \rightarrow |O_1\rangle$, where
$|O_0\rangle$ is the initial state of the observer, which
also includes her measurement apparatus, the air in the
room etc., and $|O_1\rangle$ is the final state of the
observer corresponding to the conscious experience of
having found the measurement result ``$|S_1\rangle$ !''.
Note that in our notation the measured quantum system has
become part of the extended observer state $|O_1\rangle$
(if it was a photon, for example, it may have been absorbed
by a detector). There is no assumption that the measurement
can be repeated with the same system \cite{zurekortho}.

Now suppose that the same measurement procedure, when fed
the system state $|S_2\rangle$, always gives the definite
result ``$|S_2\rangle$ !''. This corresponds to a different
state of experience for the observer, which we denote
$|O_2\rangle$, such that the dynamics of the measurement
process is now $|S_2\rangle |O_0\rangle \rightarrow
|O_2\rangle$. From our basic assumption about states of
conscious experience it follows that $\langle
O_1|O_2\rangle=0$. If the overall dynamics is unitary,
which is of course assumed in the many worlds
interpretation, then the mappings $|S_1\rangle|O_0\rangle
\rightarrow |O_1\rangle$ and $|S_2\rangle |O_0\rangle
\rightarrow |O_2\rangle$ are compatible only if $\langle
S_1|S_2\rangle=0$. Thus different possible results of an
ideal measurement always correspond to mutually orthogonal
states for the system that is being measured. The preceding
argument is inspired by, but different from, the
argumentation in Ref. \cite{zurekortho}.

Second, there is now a plausible explanation for why
observers experience definite and probabilistic outcomes in
typical quantum measurements. If the quantum system is
originally in the state $\alpha |S_1\rangle + \beta
|S_2\rangle$ (with $\langle S_1|S_2 \rangle=0$), then,
according to the above dynamics, at the end of the
measurement process we will have a state
$|\psi\rangle=\alpha |O_1\rangle + \beta |O_2\rangle$.
However, according to our above arguments, only the states
$|O_1\rangle$ and $|O_2\rangle$ correspond to definite
experiences. There is no single state of experience
corresponding to the superposition state. We thus have to
conclude that, from the point of view of the observer, the
state $|\psi\rangle$ is a superposition state of two
different ``worlds''. In one world, the observer
experiences outcome 1, in the other one outcome 2.

Since the observer always experiences definite outcomes,
she will, on repeating a given experiment, find outcome 1 a
certain number of times, outcome 2 a certain number of
times etc. (in general there will be more than two possible
outcomes). The observer will therefore be led to develop a
probabilistic description. Can we understand, based only on
the assumption of global unitary dynamics, why the observer
will be led to the Born rule? I will now argue that, third,
we can at least make a very significant step in that
direction \cite{born}.

The observer can of course measure the system in different
(system) bases. Her goal is thus to assign probabilities to
all these different bases. But Gleason's theorem
\cite{gleason} tells us that there is only one way of
assigning probabilities to all elements of all bases in a
consistent way, namely to apply Born's rule. In the given
framework of global unitary dynamics, Born's rule is
therefore the only probability law that the observer's
experiments can possibly lead her to.

There is one qualification to this statement. Gleason's
theorem assumes that the probabilities are assigned in a
{\it non-contextual} way. That is, given a quantum state
$|\phi\rangle$ which can be written as $|\phi\rangle=\alpha
|1\rangle + \beta |2\rangle + \gamma |3\rangle$ in the
basis $\{|1\rangle, |2\rangle, |3\rangle\}$, and which can
also be written $|\phi\rangle=\alpha |1\rangle + \beta'
|2'\rangle + \gamma'|3'\rangle$, choosing a different basis
$\{|2'\rangle,|3'\rangle\}$ in the subspace spanned by
$\{|2\rangle,|3\rangle\}$, it is assumed that the
probability assigned to $|1\rangle$ for the state
$|\phi\rangle$ does not depend on the choice of basis in
the complementary sub-space, i.e. it only depends on the
coefficient $\alpha$. It is interesting to note that
standard quantum physics, i.e. the Born rule, is
non-contextual in assigning probabilities, while
non-contextual ``hidden value'' assignments are of course
impossible \cite{kochenspecker}.

Is it satisfactory to have non-contextuality of the
probability assignment as an assumption? Phrased as above
(the probability only depends on the coefficient $\alpha$)
it certainly seems extremely natural. However, there may
still be room for a deeper understanding. I find the
example of decoherence encouraging in this respect. The
deeper physical understanding of its workings that was
gained since 1957 has strengthened the many-worlds
interpretation very significantly. Maybe a similar further
step will eliminate all remaining doubts concerning the
inevitability of the Born rule in the many-worlds
framework.

I have argued that the key features of our experience
(measurement results corresponding to bases, definite
results occurring probabilistically, to a large extent also
the Born rule) emerge naturally in the many-worlds
framework if we place ourselves, i.e. conscious observers,
at the center of the consideration. Of course this does not
mean that quantum physics is universally valid. This will
be decided, and quite possibly proven not be the case, by
future experiments, see Ref. \cite{marshall} for one such
attempt. However, I believe that the present considerations
strengthen the case for taking the many-worlds point of
view seriously, as possibly the most economical version of
quantum physics.

{\it Acknowledgements.} I would like to thank L. Vaidman
for convincing me years ago to take the many-worlds
interpretation seriously, and A. Kent, G. Kurizki and W.
Zurek for useful discussions.


\end{document}